\documentclass{icrc2009}

\usepackage{graphicx}   
\usepackage[caption=false]{caption}    
\usepackage[font=footnotesize]{subfig} 
\usepackage{fixltx2e}
\usepackage{url}

\newcommand{\shorttitle}[1]%
{\markboth{Proceedings of the 31\MakeLowercase{$^{st}$} ICRC, {\L}\'{o}d\'{z} 2009}{#1} }
\newcommand{\etal}{\MakeLowercase{\textit{et al. }}} 

\begin{document}
\title{VERITAS Discovery of Variability in the VHE $\gamma$-Ray Emission of 1ES 1218+304 }

\author{\IEEEauthorblockN{Asif Imran\IEEEauthorrefmark{1} for the VERITAS Collaboration\IEEEauthorrefmark{2}}
                           \\
\IEEEauthorblockA{\IEEEauthorrefmark{1}Department of Physics \& Astronomy, Iowa State University, Ames, IA, 50014, USA (imranisu@iastate.edu)}
\IEEEauthorblockA{\IEEEauthorrefmark{2}see R. Ong et al. (these proceedings) for a full author list or http://veritas.sao.arizona.edu/conferences/authors?icrc2009}}

\shorttitle{Imran \etal VERITAS 1ES1218+304}
\maketitle

\begin{abstract}

 1ES~1218+304 is among a group  of several TeV blazars that exhibit surprisingly hard energy spectra given their redshifts. The VERITAS collaboration has carried out an intensive observing campaign of 1ES 1218+304 in early 2009 to characterize the emission properties of one of these hard-spectra distant blazars in more detail.  While hard-spectra TeV blazars are most promising for unveiling putative signatures from EBL absorption, they may also have implications on shock acceleration scenarios in relativistic jets.  Our observations of 1ES~1218+304 demonstrate significant flux variability.  Furthermore, we present an energy spectrum extending into the multi-TeV regime.  The implications of these data for $\gamma$-ray emission models are discussed.
 \end{abstract}


\section{Introduction}
With over 20 confirmed TeV blazars (see~\cite{weekes}), TeV $\gamma$-ray astronomy has seen an explosion in the number of sources over recent years. Blazars are crucial to our understanding of the EBL since TeV photons are known to be attenuated by the ambient photons via the $\gamma\gamma \rightarrow e^{+}e^{-}$ interaction~\cite{gould}\cite{stecker}. Several of the newly discovered blazars at relatively large redshifts are observed to have hard energy spectra\footnote{1ES~1101-232~\cite{1es1101}, 1ES~0229+200~\cite{1es0229}, 1ES 1011+496~\cite{1es1011}, 1ES 1218+304~\cite{magic1218, pascal1}, 1ES~0347-121~\cite{1es0347}, RGB J0710+591~\cite{rgbj0710, rgbj0710-2}}. Lower limits from galaxy counts~\cite{levenson08} suggest that the source spectra of these distant blazars are surprisingly hard after correcting for EBL absorption~\cite{krenn2008}. Consequently, models of VHE emission from TeV blazars that involve standard shock acceleration and synchrotron-self-Compton (SSC) scenarios are being tested if not challenged by the presence of such hard intrinsic spectra~\cite{markus}.

 \begin{figure}[!t]
 \centering
 \includegraphics[width=3.0in]{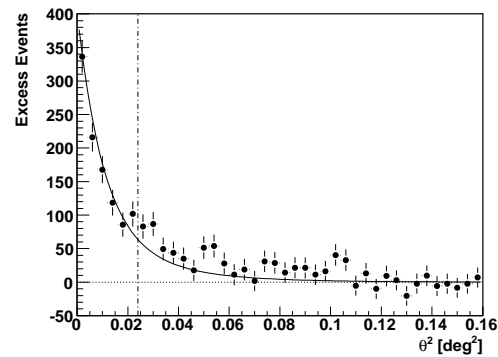}
 \caption{The distribution of $\theta^{2}$ for excess events $(ON-OFF)$ from the observation of 1ES 1218+304. The dashed-dotted line shows the boundary of the signal region ($\theta=0.154^{\circ}$). The solid curve indicates the expected  $\theta^{2}$ distribution from a point source. }
 \label{t2_fig}
 \end{figure}

 \begin{table*}[th]
 \caption{Summary of observations and analysis of 1ES 1218+304. Total live-time, range of zenith angles for observations, significance of the excess events. Also shown, are the integral flux above 200 GeV assuming a spectral index, $\Gamma$, of $\Gamma=3.18$ and the corresponding flux in units of integrated Crab Nebula flux over 200 GeV.}
 \label{tb_result}
 \centering
 \begin{tabular}{cccccc}
 \hline
 \hline \\
    &  Live Time  & Zenith      & Significance & $\Phi({\rm{>200 GeV}})$  & Crab \\
    &      [hours]   & [$^{\circ}$] & [$\sigma$]   & [$10^{-8}~m^{-2}s^{-1}$] &unit \\
    &&&&& \\
  \hline \\
   2006-2007\cite{pascal1} &17.4 & 2--35 & 10.4   & $12.2\pm2.60_{stat}$  & $0.05\pm0.011$  \\
    &&&&&\\
   2008-2009               & 30  & 2--30 & 23.5  & $17.4\pm0.89_{stat}$ & $0.07\pm0.004$ \\
 \hline
 \end{tabular}
 \end{table*}

In order to better understand the  emission characteristics of these hard spectrum blazars, the VERITAS collaboration has carried out an intensive observing campaign of the TeV blazar 1ES1218+304 in the Spring of 2009. TeV $\gamma$-ray emission from 1ES 1218+304 (redshift $z=0.182$),~a high-frequency peaked BL Lac (HBL) object, was initially discovered by the MAGIC collaboration in 2005~\cite{magic1218} and soon confirmed by VERITAS in 2006~\cite{pascal1}. Previous observations of 1ES 1218+304 by VERITAS~\cite{pascal1} resulted in a strong detection and an integral flux measurement of $\Phi(>$200 GeV$)= 12.2 \times 10^{-8}~{\rm{photons~m^{-2}s^{-1}}}$ (corresponding to $\sim$ 5\% of the flux of the Crab Nebula).  \\

\section{Observations}
The VERITAS observatory is an array of four 12 meter imaging atmospheric Cherenkov telescopes (IACT), located at the F. L. Whipple Observatory in southern Arizona~\cite{weekesver}. Each telescope is fitted with 350 hexagonal mirrors following the Davies-Cotton design. The focal plane instrument for each telescope consists of 499 fast photomultiplier tubes and preamps spanning a 3.5$^{\circ}$ field of view. The stereoscopic system combined with a large collection area ($\rm \sim10^{5}~m^{2}$) and a wide field of view allows VERITAS to be sensitive over a range of energies from 100 GeV to 30 TeV. At present, the energy resolution of single $\gamma$-rays is 15-20\% and the telescope array has an angular resolution of $\rm \sim 0.1^{\circ}$. For further details about VERITAS, see e.g. \cite{ongveritas}.

1ES~1218+304 was observed with VERITAS  for a total of 39.2 hours from December 2008 through May 2009. Data taken during bad weather, with unstable trigger rates or other anomalies were selectively removed from the data set. These quality selection cuts yielded a final exposure time of 30 hours at a mean zenith angle (ZA), of  $\rm ZA$=12$^{\circ}$. All observed data were taken in $\it{wobble}$ mode where the source is offset by $\rm 0.5^{\circ}$ relative to the center of the field of view (FoV). This allows us to simultaneously sample background events, utilizing part of the camera that is identically offset from the center of the FoV but away from the source.\\

\begin{figure}[!t]
 \centering
 \includegraphics[width=3.0in]{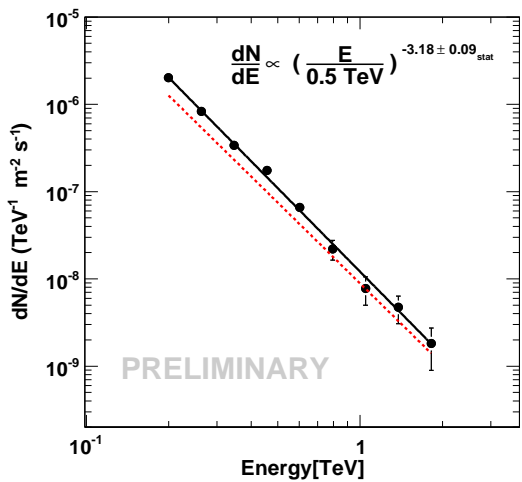}
 \caption{The observed differential energy spectrum of 1ES 1218+304 from the 2008-09 data set (preliminary). Solid line shows the best fit power law to the data (filled circles) with a spectral index value of $-3.18\pm0.09_{stat}$ . This agrees, within statistical and systematic uncertainties, to the previous VERITAS measurement of $-3.08\pm0.34_{stat}$ (dashed line)~\cite{pascal1}.}
 \label{spectrum_fig}
 \end{figure}

\begin{figure}[!t]
 \begin{center}
   \includegraphics[width=3.0in]{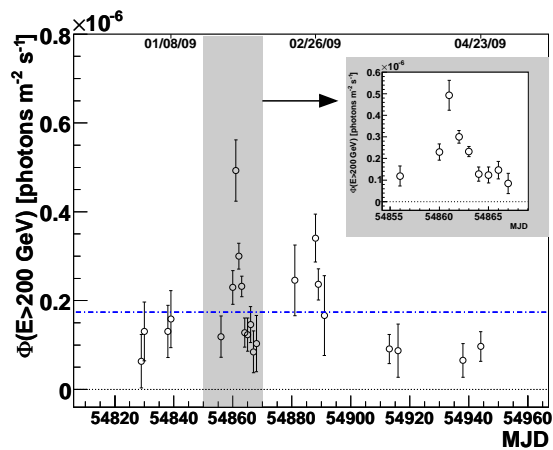}
 \end{center}
 \caption {The daily light curve in units of integral flux above 200 GeV, $\phi({\rm{E>200 GeV}})$ from 1ES 1218+304 assuming a spectral shape, dN/dE $\propto$ E$^{-\Gamma}$ with $\Gamma = 3.18$. The inset shows the flaring nights in more detail.}
 \label{lcspring_fig}
\end{figure}

\begin{figure*}[th]
 \centering
 \includegraphics[width=5.95in]{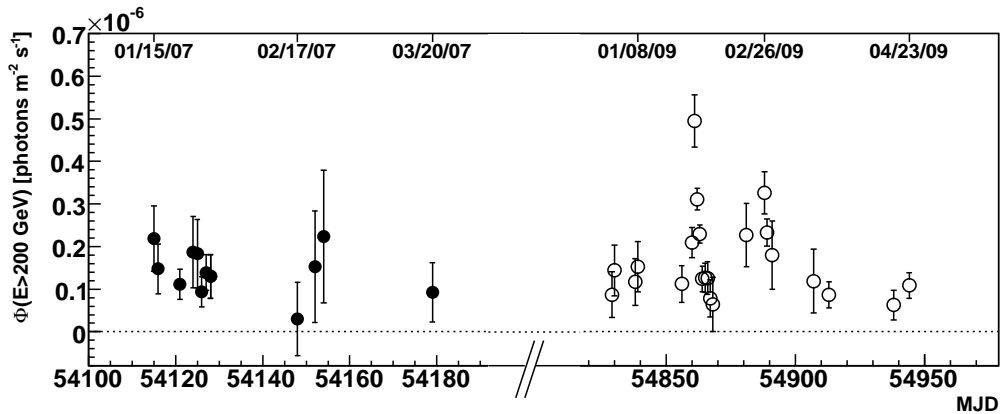}
 \vspace{-0.1in}
 \caption{The yearly light curve in units of integral flux above 200 GeV, $\phi({\rm{E>200GeV}})$ from 1ES 1218+304 is shown (binned by day). The filled circles correspond to VERITAS data from the 2006-07 epoch. The open circles represents the most recent data from the 2008-09 season.
   }
 \label{lcyearly_fig}
 \end{figure*}

\section{Data Analysis }
Following calibration of the camera gains with nightly laser runs, the data were processed with several independent analysis packages. During the analysis process, the shower images are cleaned and characterized with the standard Hillas parameters~\cite{hillas85} by computing various  moments of the light distributions. Subsequently, stereo images from different telescopes are combined to reconstruct the arrival directions on the sky and the impact point on the ground. Gamma-ray like events are identified using cuts on the {\it{mean-reduced-scaled length}} and {\it{mean-reduced-scaled width}} parameters (see ~\cite{benbow} for a description of the parameters). The energy of reconstructed events is estimated using Monte Carlo simulations. A circular source region of radius $\theta=0.154^{\circ}$ centered around the direction of 1ES 1218+304 is used to calculate the ON counts. The background or OFF counts are estimated with the $\it{Reflected-Region}$~\cite{ref_region} model using seven off-source regions. The set of $\gamma$/hadron separation cuts for this work was optimized for a source with a 10\% Crab-like flux. The significance of any excess $(ON-OFF)$ events is calculated following the method of Eq. (17) in  Li \&  Ma~\cite{lima}.\\

\section{ Results }
The result of VERITAS observation and analysis of 1ES 1218+304 in 2008-09 is summarized in Table \ref{tb_result} along with the results from 2006-07. A significant excess of 23.5$\sigma$ is detected from the direction 1ES 1218+304. Figure~\ref{t2_fig} shows the distribution of the squared angular distance, $\theta^{2}$, between the reconstructed and source position for any excess events $(ON-OFF)$. The distribution of excess events (Fig.~\ref{t2_fig}) clearly peaks at small values of $\theta^{2}$, consistent with the detection of a point-like source.

The time-averaged differential energy spectrum observed from 1ES 1218+304 during the 2008-09 observing season is shown in Fig.~\ref{spectrum_fig}. The energy spectrum ranges from 200 GeV to 1.8 TeV. The best power law fit to the measured spectral points yields,

\begin{equation}
 \frac{dN}{dE} \propto (\frac{E}{\rm{0.5TeV}})^{-3.18\pm0.09_{stat}}
 \label{power_eqn}
\end{equation}
where the flux normalization constant at 500 GeV is  $(10.94\pm0.68)\cdot10^{-8}~{\rm{TeV^{-1}m^{-2}s^{-1}}}$ with a $\chi^{2}$ of 3.4 for 7 degrees of freedom.

Figure~\ref{lcspring_fig} shows the daily integrated photon flux above 200 GeV for the 2008-09 observing season. The photon flux was calculated assuming a differential energy spectrum of dN/dE $\propto$ E$^{-\Gamma}$ with $\Gamma$=3.18. The average integral photon flux above 200 GeV is found to be (17.4$\pm0.89_{stat})\times~10^{-8}~\rm{m^{-2}s^{-1}}$ corresponding to $\sim$7\% of the Crab Nebula's flux. Strong nightly flux variations are evident in the daily light curve of 1ES 1218+304 between January 25 and February 5, 2009. During the highest state, the $\gamma$-ray flux reached $\sim$20\% of the Crab Nebula flux.  The chi-squared probability of constant emission during the entire season (Fig.~\ref{lcspring_fig}- dashed-dotted line) is $2\cdot10^{-16}$. The inset within Fig.~\ref{lcspring_fig} depicts the flaring activity in greater detail. The highest flux seen from 1ES 1218+304 occurred on MJD 54861 (January 30, 2009). The high state data in January of 2009 indicates variability (flux doubling) on a time scale of days.

Figure~\ref{lcyearly_fig} shows the daily integral photon flux for both the 2006-07 and 2008-09 seasons. We note that, except for the flaring states, the $\gamma$-ray emission is relatively constant over both periods of observation.

\section{Discussion \& Conclusions}
 The VERITAS observation of 1ES 1218+304 shows that the observed energy spectrum is considerably hard given its redshift. Moreover, the light curve clearly shows evidence of strong flux variability in the VHE gamma-ray emission of 1ES 1218+304.

B\"ottcher et al.~\cite{markus} recently suggested a possible explanation for the hardness of the VHE spectra of high redshift $\gamma$-ray blazars. According to this model, TeV blazars may contain a separate but very hard, slowly varying emission component beyond several hundred GeV. In the presence of a low magnetic field, electrons in the extended kiloparsec scale jets of TeV blazars may inverse-Compton scatter ambient cosmic microwave background (CMB) photons to TeV energies. Given the required radiative cooling time scale of the process, this TeV inverse-Compton emission is predicted to be very slowly variable in a low B-field.

Considering the time scale of the observed flux variability from 1ES 1218+304, we can constrain the size of the emission region, R, of the VHE photons to $R\leq c\cdot\Delta t\cdot\delta \approx10^{-2} \delta$~pc, where the $\delta$ is the relativistic Doppler factor of the radiating region. For a resonable value of the Doppler factor ($\leq$90), $R \ll kpc$. Therefore, we can rule out the CMB-Inverse-Compton model of emission in the kpc scale extended jets as the sole explanation for extreme hardness in the intrinsic spectra of TeV blazars.


\section*{Acknowledgments}
This research is supported by grants from the U.S. Department of
Energy, the U.S. National Science Foundation, the Smithsonian
Institution, by NSERC in Canada, by STFC in the U.K. and by Science Foundation Ireland. We acknowledge the excellent work of the technical support staff at the FLWO and the collaborating institutions in the construction and operation of the instrument.

\end{document}